\documentclass[aps,prd,preprint,groupedaddress]{revtex4}

\usepackage{amsmath,amsfonts,amssymb,bm}
\usepackage{graphicx} %

\usepackage[hang,normalsize,bf]{subfigure} %

\def\Pom{I\!P}

\begin{document}

\title{Perturbative QCD \\
 versus\\
 pion exchange and hadronic FSI effects \\
 in the $\gamma \gamma \rightarrow \pi^+ \pi^-$ reaction}

\author{A.~Szczurek}
\affiliation{Institute of Nuclear Physics, PL-31-342 Cracow, Poland}
\affiliation{University of Rzesz\'ow, PL-35-959 Rzesz\'ow, Poland}
\email{Antoni.Szczurek@ifj.edu.pl}
\author{J.~Speth}
\affiliation{Institut f\"ur Kernphysik, Forschungszentrum J\"ulich,
D-52425 J\"ulich, Germany}      

\date{\today}

\begin{abstract}
The interplay of pQCD, pion exchange and FSI effects
is studied for the
$\gamma \gamma \rightarrow \pi^+ \pi^-$ reaction in the region
of 2 GeV $< W_{\gamma \gamma} <$ 6 GeV.
We find strong interference effects between pQCD and soft
pion-exchange amplitudes up to $W_{\gamma \gamma} \sim$ 4 GeV.
We discuss to which extend the conventional hadronic FSI effects
could cloud the pQCD effects. We study multipole soft and hard
scattering effects as well as the coupling between final state
hadronic channels. We show how the perturbative effects in
$\gamma \gamma \rightarrow \rho \rho$ may mix with perturbative
effects in $\gamma \gamma \rightarrow \pi^+ \pi^-$.
The effects discussed in this paper improve
the agreement with the new data of
the DELPHI and ALEPH collaborations.
We give estimates of the onset of the pQCD regime.
Predictions for $\gamma \gamma \rightarrow \pi^0 \pi^0$
are presented.
\end{abstract}


\maketitle

\section{Introduction}

It was predicted long ago \cite{BL81} that at large $t$ and large $u$
the angular distribution of pions from the reaction
$\gamma \gamma  \rightarrow \pi^+ \pi^-$ should be described
by means of perturbative QCD as due to the exchange of t-channel (u-channel)
quarks. This reaction is commonly consider as a gold-plated
reaction for pQCD effects to be observed.

In order to have both $t$ and $u$ large,
large entrance photon-photon energies $W_{\gamma \gamma}$ are required.
Actually up to now there is no common consensus
how large $t$ (or $u$) should be so that pQCD behaviour
of the angular distributions could be observed.
At present the reaction can be studied by means of bremsstrahlung
of two photons at $e^+ e^-$ colliders only. This method is efficient
only low $W_{\gamma \gamma}$ which can be easily understood in the
equivalent photon approximation (EPA). Furthermore the pQCD predicts
a strong decrease in the cross section with increasing
photon-photon energy $W_{\gamma \gamma}$.
Thus it becomes clear that the bremsstrahlung method can be efficiently
used only for not too high $W_{\gamma \gamma}$ energies.

In the leading-twist perturbative treatment, the transition amplitude
factorizes into a hard scattering amplitude and pion distribution
amplitude. The pion distribution amplitude was found to be
strongly constrained by the photon-pion transition form factor
\cite{transition_formfactor}.
For the "realistic" distribution amplitude the pQCD contribution
to the $\gamma \gamma \rightarrow \pi^+ \pi^-$ \cite{Vogt}
is well below the experimental data above the resonance region
( 2 GeV $<  W <$ 3.5 GeV ) \cite{TPC,CLEO}.
Recent analyses of the DELPHI and ALEPH collaborations at LEP2
 \cite{DELPHI,ALEPH} extend the energy range.
In contrast to previous measurements it was possible to
separate $\pi^+ \pi^-$ and $K^+ K^-$ channels in the recent analyses.
The preliminary data confirm the missing strength problem.

Thus it becomes clear that the leading-twist pQCD approach is
not sufficient to describe the experimental data and
the mechanism must be of nonperturbative origin.
In a recent work \cite{DKV02} it was proposed that the
amplitude for $\gamma \gamma \rightarrow \pi^+ \pi^-$ to be described
as a hard $\gamma \gamma \rightarrow q \bar q$ amplitude
times a form factor describing the soft (not calculable)
transition from $q \bar q$ to the meson pair.
It was assumed implicitly in this approach that all the
strength comes from the so-called hand-bag process !

In the present study we take slightly different approach
in an atempt to establish to what extend the present data can
be understood in terms of relatively well known soft hadronic physics
and what is the role of final state interaction (FSI) effects
not yet discussed in the literature.
At energies below $f_2$ resonance the FSI effects are known
to be rather weak \cite{MP87} and the cross section is dominated
by one-pion exchange. Above the $f_2$ resonance the FSI effects
have never been studied.
We have shown recently \cite{SNS2001} that in the pion-pion
elastic scattering at $W >$ 2 GeV the multipole soft rescattering
may play an important role up to very large momenta transferred, i.e.
it shifts the onset of the pQCD behaviour to larger $t$ or $u$.

Inspired by the result of \cite{SNS2001}, in the present note
we discuss the role of the final state interaction effects
in $\gamma \gamma \rightarrow \pi^+ \pi^-$. We shall show that
at not too high energies, i.e. where experiments were performed,
the coupling between the $\rho \rho$ and the $\pi \pi$ channels
may play quite important role.

\newpage

\section{Perturbative QCD approach and \\ corrections in the soft regions}

It is well known that at large momentum transfers the exclusive
processes, such as $\gamma \gamma \rightarrow \pi^+ \pi^-$,
test Quantum Chromodynamics \cite{pQCD}.
It is not known, however, how big the momentum transfer
should be in order that the rules of perturbative QCD (pQCD)
could be applied.

According to the rules proposed in \cite{pQCD},
at large energies and large center-of-mass angles, i.e.
at large four-momentum transfers, the amplitude
for the $\gamma \gamma \rightarrow \pi^+ \pi^-$ can be
factorized into a perturbatively calculable hard-scattering amplitude
and a non-perturbative distribution amplitude of finding a valence quark
in each pion
\begin{equation}
M_{\lambda_1 \lambda_2}(W,\theta) =
\int_0^1 dx_1 \int_0^1 dx_2 \; \Phi_{\pi}^*(x_1, \tilde{Q}_1) \;
                        T_{\lambda_1 \lambda_2}^{H}(x_1, x_2, W, \theta) \;
                               \Phi_{\pi}^*(x_2, \tilde{Q}_2) \; .
\label{pQCD_amplitude}
\end{equation}
The indieces $\lambda_1$ and $\lambda_2$ are photon helicities.
In general the distribution amplitudes $\Phi_{\pi}$ undergo a slow logarithmic
QCD evolution, depending on scales ${\tilde Q}_1$ and ${\tilde Q}_2$.
In the collinear approach $T^H$ is computed assuming that the
quarks (antiquarks) are collinear with outgoing mesons.
Because in the present analysis we concentrate mainly on
pion exchange and FSI effects,
we shall limit the approximatiom for $T^H$ to the leading order (LO)
in $\alpha_s$ only.
\footnote{The next to leading order (NLO) calculation \cite{NLO} leads
to a reduction of the LO result.
Therefore the LO result can be viewed as an upper estimate of pQCD.}
In practice the distribution amplitudes cannot be calculated
from first principles. Based on phenomenology of a few reactions,
it seems at present that the distribution amplitude should be rather
close to the asymptotic one:
\begin{equation}
\Phi_{\pi}(x) = \sqrt{3} f_{\pi} x (1-x) \; ,
\label{asymptotic_distribution_amplitude}
\end{equation}
where $f_{\pi}$ is the pion decay constant ($\sqrt{2} f_{\pi}$ =
130.7 MeV).

\begin{figure}
\includegraphics[width=9cm]{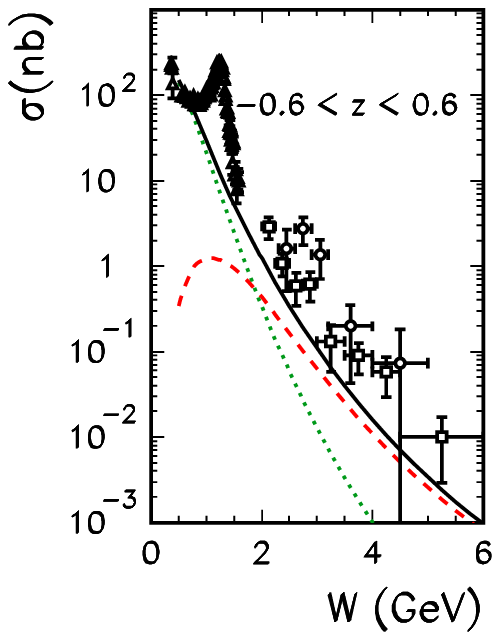}
\caption{
  Predictions of the collinear LO pQCD (dashed line) for the cross
  section of the $\gamma \gamma \rightarrow \pi^+ \pi^-$ reaction
  integrated in the experimental range of $-z_{exp} < cos\theta <
  z_{exp}$ versus experimental data from the MARKII collaboration
  \cite{Mark2} (triangles), the DELPHI collaboration \cite{DELPHI}
  (circles), and the ALEPH collaboration \cite{ALEPH} (squares).  The
  dotted line corresponds to the contribution of the pion-exchange
  mechanism discussed in the text.  The coherent sum of the pQCD and
  pion-exchange contribution is shown by the solid line.
\label{fig_pQCD}}
\end{figure}

In Fig.\ref{fig_pQCD} we compare the pQCD result obtained for the
asymptotic distribution with
the experimental data from \cite{DELPHI,ALEPH}.
In our calculation we follow \cite{JA90} rather
than \cite{BL81} and use running coupling constant with
simple analytic models for freezing $\alpha_s$ \cite{JA90,SS97}.
There is a clear excess of the experimental cross section
over the pQCD prediction in the whole measured range of
photon-photon energies. What other processes can contribute
in this energy range?
Can the pion exchange and/or final state interaction be responsible
for the missing strenght?
We shall try to answer these questions in the following sections.

The pQCD amplitude (\ref{pQCD_amplitude}) contains
singularities at $\theta$ = 0 and $\theta$ = $\pi$ (see \cite{BL81})
which are artifacts of the collinear approximation.
This is the region of the phase space where the collinear pQCD result
is certainly not trustworthy. Furthermore the small angle scattering is
probably of soft nonperturbative character due to meson or reggeon
exchanges
which are included in the present approach explicitly.
Therefore in order to avoid double counting and make possible
the use of the pQCD amplitudes in the multiple-scattering series,
we shall smoothly cut off the small $t$ and small $u$ regions
in the perturbative amplitude. This is done phenomenologically
by introducing a $t-u$ symmetric form factor
\begin{equation}
F_{reg}^{pQCD}(t,u) =
\left[1 - \exp\left(\frac{t-t_m}{\Lambda_{reg}^2}\right) \right] \cdot
\left[1 - \exp\left(\frac{u-u_m}{\Lambda_{reg}^2}\right) \right] \;
\label{pQCD_regulator}
\end{equation}
which the pQCD amplitude (\ref{pQCD_amplitude}) is multiplied by.
Here, $t_m$ and $u_m$ are maximal kinematically allowed values
of $t$ and $u$, respectively,
and $\Lambda_{reg}$ is a cut-off parameter to be fixed.
The above choice for the form factor removes the singularities in
the original collinear pQCD amplitude.
Our model division
into soft and hard regions suggests that $\Lambda_{reg}$ should be of
the order of the cut off parameter in the vertex form factor for
soft meson/reggeon exchanges, i.e. of the order of 1 GeV.
\footnote{Our cut off parameter $\Lambda_{reg}$ can be also conveniently
used to simulate effects of finite transverse momenta of quarks
in the final pions which leads to a suppression of collinear
pQCD result up to s$\sim$20 GeV$^2$ \cite{Vogt}. In order to
simulate phenomenologically the results of \cite{Vogt} we would need
also $\Lambda_{reg} \approx$ 1 GeV.}

\begin{figure}
\includegraphics[width=9cm]{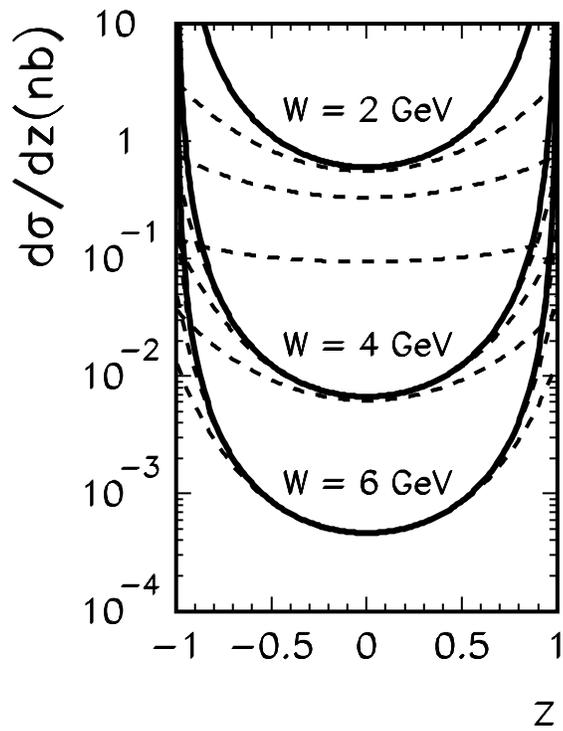}
\caption{
  The pQCD contribution to the angular distribution for the $\gamma
  \gamma \rightarrow \pi^+ \pi^-$ reaction for W = 2, 4, 6 GeV.  The
  result of the standard calculation is shown by the solid lines.  The
  dashed line represent the results obtained when the hard scattering
  amplitude was modified by the form factor (\ref{pQCD_regulator})
  correspondingly for three different values of $\Lambda_{reg}$ = 0.5,
  1.0, 2.0 GeV.
\label{fig_pqcd_cut}}
\end{figure}

In Fig.\ref{fig_pqcd_cut} we show $d \sigma / dz$ ($z=\cos(\theta)$)
obtained from the pQCD amplitude (\ref{pQCD_amplitude})
modified by the form factor (\ref{pQCD_regulator}) for different
values of $\Lambda_{reg}$ = 0.5, 1.0, 2.0 GeV,
for three different energies W = 2,4,6 GeV.
In the z-distribution the soft/hard "borders"
are clearly energy dependent which can be seen by deviations of
the modified pQCD results (dashed lines) from the original pQCD
results (solid lines).
At low energies dominant mechanism becomes nonperturbative.
When $s \rightarrow 4 m_{\pi}^2$ the perturbative component is
totally suppressed by $F_{reg}^{pQCD}(t,u) \rightarrow$ 0.
Then, only soft processes known from low and intermediate energy
physics are at play. Thus in the low energy limit our approach
coincides with the succesful low-energy phenomenology by construction.
Only at high energies one can expect pQCD effects to dominate
in the region of intermediate $z$ ($\theta \approx \pi/2$).


\newpage

\section{QED pion exchange \\ and finite-size corrections}

The QED Born amplitude for the $\gamma \gamma \rightarrow \pi^+ \pi^-$
reaction with point-like particles has been known for long time
\cite{pion_exchange}.
At low energies, where chiral perturbation theory is usually applied,
pions are treated as point-like particles.
The finite-size corrections vanish automatically
when $s \rightarrow$ 0, i.e. $t,u \rightarrow$ 0 as well.
This means that only at sufficiently high energies
the finite-size corrections may show up.
 
An interesting problem is how to generalize the QED amplitude
for real finite-size pions.
For the sake of simplicity we follow the idea of Poppe \cite{Poppe}
and correct the QED amplitude by an overall $t$ and $u$
dependent form factor
\begin{equation}
A_{fs}(t,u,s) = \Omega(t,u,s) \cdot A_{QED}(t,u,s) \; .
\label{finite_size_corrections}
\end{equation} 
This form by construction guarantees crossing symmetry.
It is natural to expect that finite size corrections should damp
the QED amplitude for both $t$ and $u$ large.
At sufficiently high energies the following simple Ansatz fulfils
this requirement
\begin{equation}
\Omega(t,u,s) = \frac{F^2(t) + F^2(u)}{1 + F^2(-s)} \; ,
\label{FF_ansatz}
\end{equation}
where F are standard vertex function such that $F(0) \approx$  1
and $F(t) \rightarrow $0 when $t \rightarrow \infty$.
\footnote{In our case of light pions it is rather academic if
the form factors are normalized at t,u=0 or
on (pion) mass shell t,u = $m_{\pi}^2$.}
The Ansatz (\ref{FF_ansatz}) has a nice feature that
in the limit of large $s$
\begin{eqnarray}
&&\Omega(t,u,s) \stackrel{t \rightarrow 0}{\rightarrow} F^2(t) \; , 
\nonumber \\
&&\Omega(t,u,s) \stackrel{u \rightarrow 0}{\rightarrow} F^2(u) \; ,
\label{FF_in_limits}
\end{eqnarray}
i.e. it generates standard vertex form factors.
Then at $z \equiv \cos\theta \approx$ 0 (large $s$),
 $\Omega(t,u,s) \approx 2 F^2(-s/2)$,
i.e.
\begin{equation}
\frac{d \sigma^{\pi}}{dz}(z = 0, s) \propto F^4(-s/2) \; .
\end{equation}
For the monopole vertex form factor this leads to
a suppression $\propto 1/s^4$ of the QED result for
point-like particles. Thus the pion exchange gives a faster
decrease of the cross section with energy than the pQCD mechanisms,
and at sufficiently
high energy it can be neglected. An interesting question is:
at what energy does this difference become significant.

The correct small and large angle behaviour (\ref{FF_in_limits})
allows us
to use traditional ($t$- or $u$-dependent only) vertex form factors.
We note, however, that the exact form of $F$ is not known,
particularly in the region of interest, i.e. at large $t$ and $u$.
In order to have some flexibility when discussing the effect of the poorly
known large t/u region we choose a simple but rather flexible form
for the vertex form factors
\begin{equation}
F_t(t) = \exp \left(\frac{B_{\gamma \pi}}{4} t\right) \; , \; 
F_u(u) = \exp \left(\frac{B_{\gamma \pi}}{4} u\right) \; .
\label{gamma_pion_slope}
\end{equation}
From the analysis in \cite{SNS2001} we expect
$B_{\gamma \pi} <$ 4 GeV$^{-2}$
for the elementary vertex. 
In order to see the sensitivity to the particular functional form of $F$
we shall also replace $F$'s by the pion electromagnetic form factor $F_{EM}$,
assuming the standard monopole form with the $\rho$ meson mass.
\footnote{This seems natural as in our case $F$ can be interpreted as
half off-shell electromagnetic form factor $F(Q^2=0,m_{\pi}^2,t/u)$
of the pion, with only one pion in the vertex being off its mass shell.}

In order to complete our discussion and quantify the finite size effects
we show the result obtained for one-pion exchange model with
$F = F_{EM}$ in Fig.\ref{fig_pQCD} (dotted line).
It is surprising that the so-obtained finite-size-corrected pion exchange
is comparable with the pQCD result (dashed lines) up to relatively large
energies.
If both amplitudes are of comparable size one has to include
interference effects.

\section{The pQCD and pion exchange amplitudes interfere!}

The size of the interference effects can be estimated by
comparing the incoherent sum of pQCD process and soft pion exchange
$\sigma_{inc} \propto |A_{pQCD}| + |A_{\pi}|^2$
and the coherent sum $\sigma_{coh} \propto |A_{pQCD}+A_{\pi}|^2$
(see the thick solid line in Fig.\ref{fig_pQCD} and the third column
in Table 1).
The interference effects are generally large and positive.
The coherent sum $|pQCD + \pi|^2$ only weakly depends
on the functional form of the form factor used.
The details depend, however, on the vertex form factor.
In the case of the monopole (EM) form factor the effect of
the interference effects survives up to large energies $W \sim$ 5 GeV
when the pion-exchange contribution alone
becomes much smaller than the perturbative contribution.
In the case of the exponential form factor
($B_{\gamma \pi}$ = 3 GeV$^{-2}$) the interference effects become
negligible for $W \sim$ 4 GeV.

As can be seen from the figure the interference of the pQCD
and pion-exchange amplitudes improves considerably the agreement
with the recent DELPHI \cite{DELPHI} and ALEPH \cite{ALEPH}
data. This automatically implies that there is less room is left for
the hand-bag mechanism contribution considered in \cite{DKV02}.

\begin{figure}
\includegraphics[width=9cm]{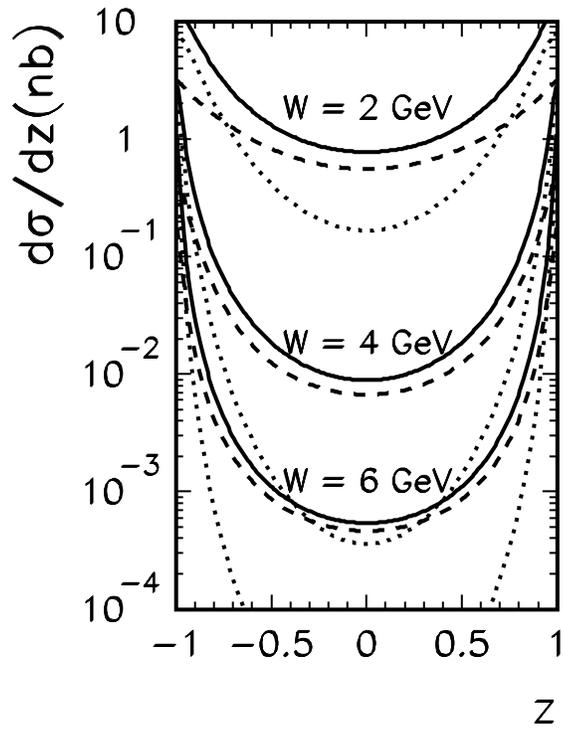}
\caption{
  Soft pion-exchange (dotted) versus hard quark-exchange mechanism
  (dashed) and their coherent sum (solid) for the $\gamma \gamma
  \rightarrow \pi^+ \pi^-$ reaction at W = 2,4,6 GeV. In this
  calculation $\Lambda_{reg}$ = 1.0 GeV and $B_{\gamma \pi}$ = 3
  GeV$^{-2}$.
\label{fig_pqcd_pion}}
\end{figure}

The effect of interference of the pion-exchange and pQCD contributions
on the angular distribution is shown in Fig.\ref{fig_pqcd_pion}.
The constructive interference can be observed over
the whole angular range.

\section{Tail of the $f_2$ resonance}

The $f_2$ resonance is known to be strongly populated
in the $\gamma \gamma$ collisions (see for instance
\cite{Mark2,Crystal_Ball}).
It is rather broad with $\Gamma \sim$ 0.2 GeV. 
The leading twist pQCD contribution drops strongly with energy.
Therefore one should look at the interplay of
the high energy flank of $f_2$ with the strongly decreasing
continuum.
Then all kinds of energy dependence of kinematical and dynamical
origin must be included.
In the relativistic approach, the total cross section for the
resonance contribution reads 
\begin{equation}
\sigma_{\gamma \gamma \rightarrow \pi \pi}(W) =
8 \pi (2J+1) \left(\frac{M_R}{W} \right)^2 
\frac{\Gamma_{\gamma \gamma} \Gamma_{tot}(W) Br(f_2 \rightarrow  \pi^+ \pi^-)}
{ (W^2-M_R^2)^2 + M_R^2 \Gamma_{tot}^2 } \; .
\label{resonance}
\end{equation}
One usually parametrizes $\Gamma_{tot}$ as
\begin{equation}
\Gamma_{tot}(W) = \Gamma_{tot}^{0}
\left( \frac{p}{p_0}  \right)^{2l+1} {\cal F}_{dyn}(p) \; ,
\label{Gamma_tot}
\end{equation}
where ${\cal F}_{dyn}(p)$ is a function of dynamical origin
usually obtained in a simple nuclear physics inspired model
of resonances \cite{BW_book}.

The analysis of angular distributions has shown
\cite{Mark2,Crystal_Ball}
that $f_2$ is produced dominantly
in the helicity $\lambda = \pm$ 2 state, in agreement with
earlier theoretical predictions \cite{GK76}.
This means that the angular distribution of pions in the
$f_2$ frame of reference factorizes as
\begin{equation}
\frac{d \sigma}{dz}(W,\theta) =
2 \pi |Y_{22}(\theta)|^2 \cdot
\sigma_{\gamma \gamma \rightarrow \pi \pi}(W) \; .
\label{angular_distribution_f2} 
\end{equation}
This formula can be used to calculate the resonance contribution
in a limited range of $\cos \theta$, as it is usually done
in experiments.

\begin{figure}
  \subfigure[]{\label{fig_f2a}
    \includegraphics[width=4.5cm]{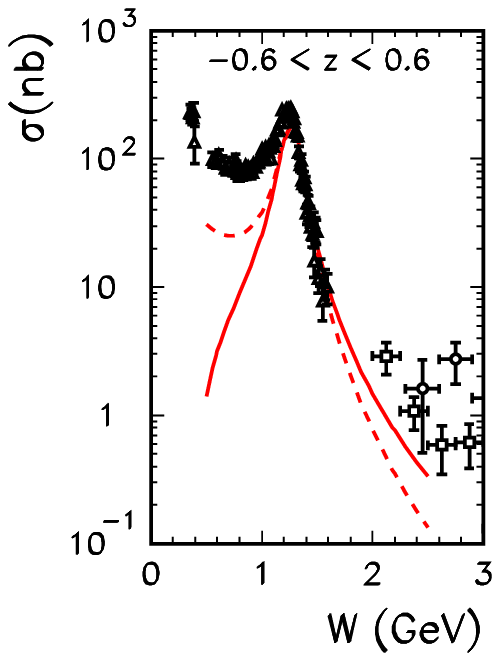}}
  \subfigure[]{\label{fig_f2b}
    \includegraphics[width=4.5cm]{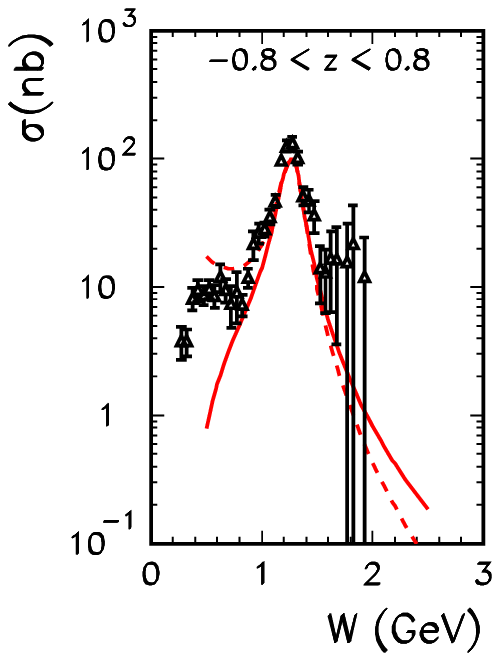}}
  \caption{
    The $f_2$ resonance contribution to the $\gamma \gamma \rightarrow
    \pi^+ \pi^-$ and to the $\gamma \gamma \rightarrow \pi^0 \pi^0$
    (left and right panel respectively).  For comparison results for
    both energy dependent (solid line) and energy independent (dashed
    line) width are shown.  The experimental data of Mark II
    \cite{Mark2} and the Crystal Ball \cite{Crystal_Ball},
    respectively, are shown for comparison.
\label{fig_f2}}
\end{figure}

In Fig.\ref{fig_f2} we show the resonance contribution
obtained from (\ref{resonance}) and (\ref{Gamma_tot}) with
other parameters taken from \cite{PDB}.
This contribution competes with both pQCD and
pion contribution up to very large energies. 
It would be naive to expect, however, that the energy dependence
of (\ref{Gamma_tot}) can be used far from the resonance region.
For comparison, we also show (dashed line) the result obtained
with the relativistic Breit-Wigner formula (\ref{resonance}) but with
an energy-independent width $\Gamma_{tot} = \Gamma_{tot}^0$.
The result with energy independent width describes slightly better
the high energy flank of $f_2$ for the $\pi^+ \pi^-$ channel. 

In the vicinity of the $f_2$ peak one may also expect interference of
the resonance amplitude and the continua due to pion exchange
and pQCD mechanisms. Because a consistent overall
microscopic model of the resonance and the continua is not available,
the relative phase between the resonance amplitude and that for
the pion exchange is not known.
In the following we take a pragmatic attitude and
try to fit the phase to the MARKII data \cite{Mark2}.
In order to explain the low energy flank of the $f_2$ resonance
we need $\phi \approx \pi/2$. We shall make the simplest approximation
and assume that this phase is energy independent.
Adding all contributions together we obtain results represented by
the thick solid
and thick dashed
lines for energy-dependent and energy-independent widths in the
resonance amplitude, respectively.
In this way we obtain a result almost consistent with the preliminary
DELPHI \cite{DELPHI} and ALEPH \cite{ALEPH} data at least up to
W = 2.5 GeV. The agreement with the data may be extended even further
towards higher photon-photon energies, but it is not clear if the
adopted simple approximations are reasonable that far from
the resonance position.

All this demonstrates that the situation just above the $f_2$
resonance is far from being under control. Therefore expecting 
the pQCD quark exchange mechanism to be the only reaction mechanism
in this region seems to us rather unjustified.

\section{FSI effects}

Up to now we have included only pion exchange Born amplitude.
It was demonstrated recently that for elastic pion-pion scattering,
soft FSI effects lead to a damping of the cross section at
small angles and a considerable enhancement in the region of
intermediate angles where they compete with the two-gluon exchange
amplitude.
Let us explore the role of FSI effects for the reaction
under consideration.

The FSI effects are an intrinsic part of fundamental scattering theory.
Therefore there should be good reasons for FSI to not occur.
Calculations of FSI effects at intermediate energy is
not always an easy task.
In the following we shall discuss only some selected FSI effects for
the reaction under consideration.


\begin{table}

\caption{\it Brief summary of double-scattering processes conidered
in this paper for $\gamma \gamma \rightarrow \pi^+ \pi^-$.
}
\begin{center}
\begin{tabular}{|c|c|c|c|c|}
\hline
number & first step & intermediate channel & second step & short notation  \\  
\hline
1 & soft $\pi$-exch. & $\pi^+\pi^-$ & $\Pom+f_2+\rho$ exchange & $S_1 S$  \\   
2 & pQCD             & $\pi^+\pi^-$ & $\Pom+f_2+\rho$ exchange & $H_1 S$  \\ 
3 & soft VDM         & $\rho^0 \rho^0$ & $\pi$ exchange       & $S_2 \pi$ \\
4 & pQCD             & $\rho^+(0) \rho^-(0)$ & $\pi$ exchange & $H_2 \pi$ \\
\hline
\end{tabular}
\end{center}
\end{table}


In
Table 2 we list 
double-scattering terms considered
in the present analysis. In order to demonstrate the role of
each mechanism separately, we shall calculate contributions of
individual diagrams.
Interference effects of different mechanisms will be discussed
only at the very end.

In the language of multiple scattering at high energies
\cite{T-M70} the amplitude for the
$\gamma \gamma \rightarrow \pi^+ \pi^-$
reaction can be written as an infinite series of the type:
\begin{eqnarray}
& & A_{\gamma \gamma \rightarrow \pi^+ \pi^-}(s,t,u) =
\sum_{\alpha} A_{\gamma \gamma \rightarrow \pi^+ \pi^-}^{(\alpha)}(s,t,u)
\nonumber \\
 &+& 
\sum_{ij} \sum_{\alpha,\beta} 
\frac{i}{32 \pi^2 s} \int d^2 \vec k_1 d^2 \vec k_2 \; 
\delta^2(\vec k - \vec k_1 - \vec k_2) \; 
 A_{\gamma \gamma \rightarrow i j}^{(\alpha)}(s,\vec k_1) \;
 A_{i j \rightarrow \pi^+ \pi^-}^{(\beta)}(s,\vec k_2)
\nonumber \\
 &+& (...) \; .
\label{multiple_scattering} 
\end{eqnarray}
Here Greek indices label type of exchange,
while Latin indices $i j$, etc. label two-body intermediate
states.

The first component in (\ref{multiple_scattering}) corresponds to
single-exchange terms.
In the following we shall include only
$\alpha = \pi$ (pion exchange) or $\alpha = 2q$ (pQCD quark exchange)
single-exchange amplitudes. Contribution of other
single-exchange processes, like $\rho, a_1, a_2$ meson/reggeon
exchanges are more difficult to estimate reliably and will
be discussed elsewhere.
We shall include only
$i j = \pi^+ \pi^-$ and $i j = \rho \rho$ intermediate states.
The latter are expected to be copiously created in the photon-photon
collisions either assuming factorization
\cite{gg_rho0rho0_factorization} or in QCD-inspired models
(see for instance \cite{GPS87}).
The parameters of pion-pion interaction are taken from Ref.\cite{SNS2001}. 

In the previous sections the uncertainty
of the pQCD and pion-exchange amplitudes has been discussed.
Now we shall discuss how reliably one can evaluate 
double-scattering amplitudes of Table 2.
In general, the double-scattering amplitudes interfere with
the single-scattering amplitudes considered before.
We shall discuss the interference pattern and the sign of
some selected interference terms.

\subsection{$\gamma \gamma \rightarrow \pi^+ \pi^-$ and FSI effects}

As already discussed there are two single exchange mechanisms
leading directly to the $\pi^+ \pi^-$ channel: soft pion exchange ($S_1$)
and pQCD hard quark exchange ($H_1$). While the first can be reliably
calculated at small $t$ or small $u$, the second is trustworthy only at
$t$ and $u$ both large. Going beyond the region of their applicability
requires extrapolations which cannot be derived from first
principles and leads to obvious uncertainties.
In particular, as discussed in the previous sections, it is not
completely clear what is the contribution of pion exchange in
the measured angular distributions for -0.6 $ < \cos\theta < $ 0.6.

In this section we shall explore the role of double-exchange terms
$S_1 S$ and $H_1 S$ where $S$ denotes soft pion-pion interaction.
The parameters of the pion-pion interaction are fixed using
the analysis of \cite{SNS2001}. 

\begin{figure}
  \subfigure[]{\label{fig_pis_soa}
    \includegraphics[width=4.5cm]{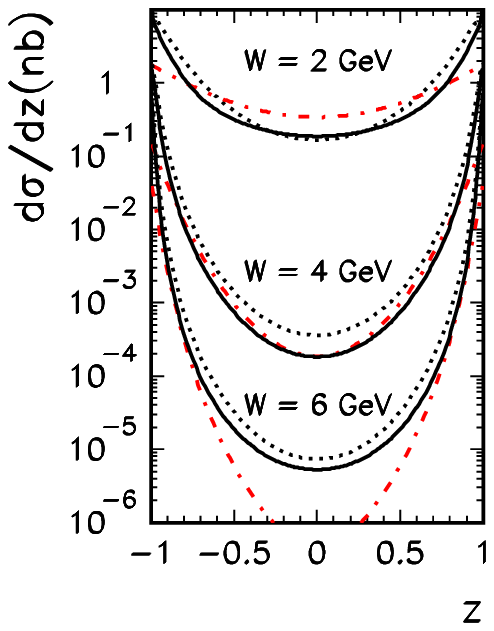}}
  \subfigure[]{\label{fig_pis_sob}
    \includegraphics[width=4.5cm]{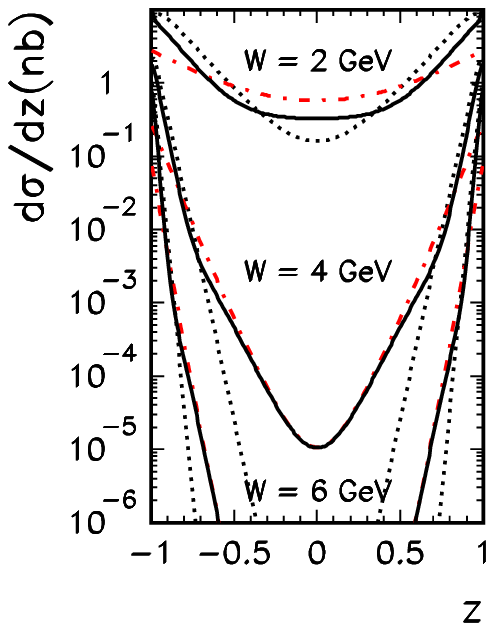}}
\caption{
  Angular distribution of pions for W = 2,4,6 GeV.  Cross section
  corresponding to double-scattering term $S_1 S$ is denoted by the
  dash-dotted line.  Left (right) panel corresponds to the monopole
  vertex form factor (the exponential form factor with $B_{\gamma
    \pi}$ = 3 GeV$^{-2}$) of the first step.  The dotted line refers
  to the Born result for pion exchange.  Their coherent sum is shown
  by the solid line.
\label{fig_pis_so}}
\end{figure}

Let us start with the $S_1 S$ term (first row in Table 2).  
In Fig.\ref{fig_pis_so} we show the cross section calculated
with the amplitude $A = S_1 S$ (dash-dotted line) with the monopole
form factors for the first step reaction (left panel) and with
the exponential form factors with $B_{\gamma \pi}$ = 3 GeV$^{-2}$
(right panel).  
The solid (dotted) line denotes
the cross section corresponding to the amplitude
$A = S_1 + S_1 S$ ($A = S_1$, i.e. pion exchange alone).
In these calculations monopole or exponential
( $B_{\gamma \pi}$ = 3 GeV$^{-2}$) form factor were used.
The answer to the question which of the amplitudes squared
$|A = \pi + S_1 S|^2$ or $|\pi|^2$ is bigger for $z \sim$ 0
depends on the half-off shell pion electromagnetic form factor.
For the exponential form factor
$|\pi + S_1 S|^2 > |\pi|^2$ while for the monopole form factor
$|\pi + S_1 S|^2 < |\pi|^2$. 
By comparing the solid and dotted lines one can infer
that the interference is destructive.
For $z \sim$ 0 the pion exchange term strongly depends
on the vertex form factor used.
It is amusing that the total result $|S_1 + S_1 S|^2$
depends on the vertex form factor up to $W \sim$ 4 GeV 
only weakly. 

\begin{figure}
\includegraphics[width=9cm]{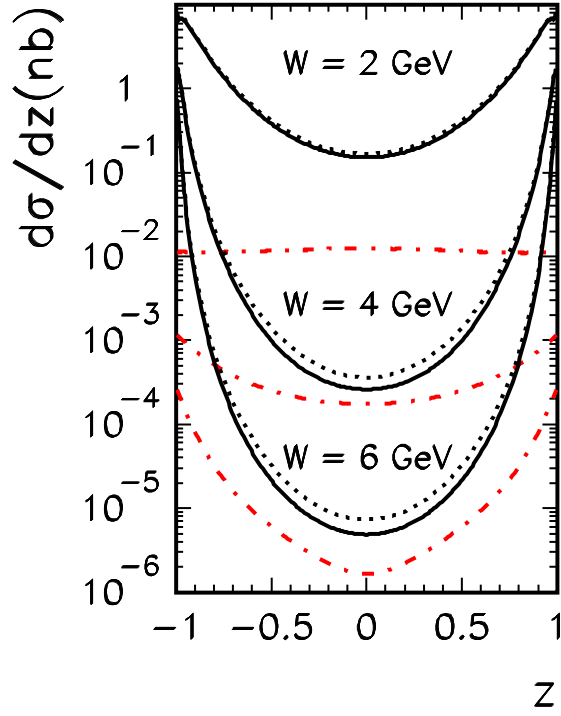}
\caption{
Angular distribution of pions for W = 2, 4, 6 GeV.
The cross section corresponding to the double-scattering
term $H_1 S$ is shown by the dash-dotted lines.
In this calculation $\Lambda_{reg}$ = 1 GeV.
For comparison also shown is the pion exchange result (dotted line).
Solid lines correspond to the coherent sum
of the $H_1 S$ and pion exchange amplitudes.
\label{fig_gg_pipih1}}
\end{figure}

Let us now turn to the $H_1 S$ double-scattering amplitude
(second row in Table 2).
As in the previous case, to check reliability of our estimates of
this double-scattering amplitude, in Fig.\ref{fig_gg_pipih1}
we present the cross section calculated with the $A = H_1 S$
amplitude alone (dash-dotted line).
In this calculation $\Lambda_{reg}$ = 1.0 GeV.
We have performed the calculations for different low-t(u) cuts
(\ref{pQCD_regulator}).
There is a relatively mild dependence on the way the soft
nonperturbative regions are treated. 
For reference we show the pion exchange result with
the monopole form factor in vertices (dotted line)
and the coherent sum of the $\pi$ and $H_1 S$ amplitudes
(solid line). In the second column of Table 3 we give
the cross section calculated with the double-scattering
amplitude integrated in the limited range of $\cos \theta$.
The hierarchy of the cross sections
can be summarized as
\begin{equation}
|H_1 S|^2 \ll |\pi + H_1 S|^2 < |\pi|^2 \; .
\end{equation}
This means that the interference effect between the pion-exchange
amplitude and the double-scattering $H_1 S$ amplitude is destructive
and reduces the cross section by 20-40 \% in the considered region of
energies and cos$\theta$.

\subsection{$\gamma \gamma \rightarrow \rho \rho$ and its coupling
to the $\pi^+ \pi^-$ channel}

For the $\rho \rho$ intermediate channel, as for the $\pi \pi$
channel before, we shall consider
both soft ($S_2$) and hard ($H_2$) mechanisms.
In both cases the second step mechanism is assumed to
be the standard pion exchange.

The usual way to calculate soft contributions to
$\gamma \gamma \rightarrow \rho \rho$ is to assume
the validity of the vector dominance (VDM) and
the factorization at the hadron level. The parameters of
the $\rho \rho$ interaction can be obtained in a similar manner
as for the pion-pion scattering in \cite{SNS2001}.
Then for $\gamma \gamma \rightarrow \rho^0 \rho^0$ we obtain:
\begin{equation} 
A_{\lambda_1 \lambda_2 \rightarrow \lambda_{\rho_1} \lambda_{\rho_2}}(t) =
\sqrt{ \frac{4 \pi}{f_{\rho}^2} }
\sqrt{ \frac{4 \pi}{f_{\rho}^2} }
\delta_{\lambda_1 \lambda_{\rho_1}}
\delta_{\lambda_2 \lambda_{\rho_2}} \;
\left[
A_{\Pom}(t) + A_f(t) 
\right]   \; ,
\label{rho0_rho0_soft}
\end{equation}
where the pomeron exchange $A_{\Pom}(t)$ and the isoscalar reggeon
exchange $A_f(t)$ amplitudes are given explicitly
by Eq.(10) in \cite{SNS2001} with parameters specified
in the text there.
\footnote{We neglect isovector $a_2$ contribution which is
more difficult to be reliably estimated.}
Above, we have additionally assumed helicity conservation.

\begin{figure}
\includegraphics[width=9cm]{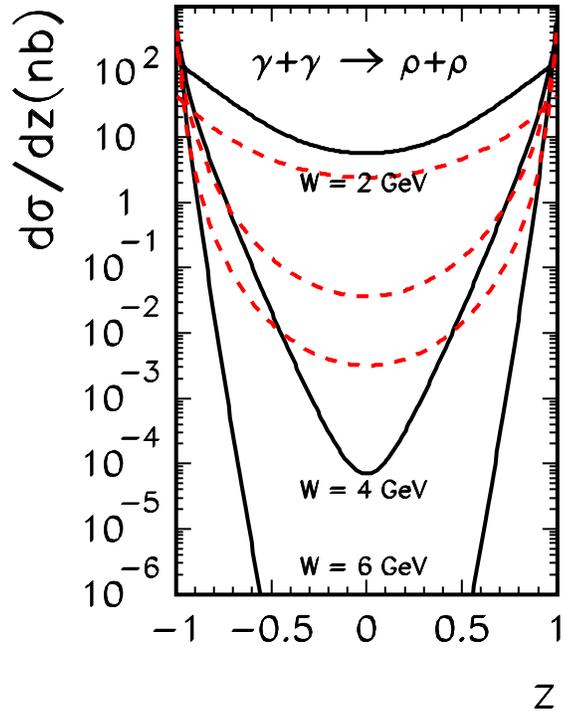}
\caption{
Angular distribution of $\rho^0$ from the reaction
$\gamma \gamma \rightarrow \rho^0 \rho^0$ (solid line)
and $\rho^{\pm}$ form the reaction
$\gamma \gamma \rightarrow \rho^+(\lambda = 0) \rho^-(\lambda = 0)$
(dashed line) for W = 2,4,6 GeV.
\label{fig_gg_rhorho}}
\end{figure}

In Fig.\ref{fig_gg_rhorho} we show
the angular distribution for the soft process
$\gamma \gamma \rightarrow \rho^0 \rho^0$.
In this calculation we have used exponential form factor with
$B_{\gamma \rho}$ = 3 GeV in the vertices of the
$\gamma \gamma \rightarrow \rho^0 \rho^0$ Born amplitude. 
In the same figure we present the result for pQCD calculation
(for the three energies $W_{\gamma \gamma}$ = 2, 4, 6 GeV)
according to \cite{BL81} with an extra form factor to cut off
soft nonperturbative regions ($\Lambda_{reg}$ = 1 GeV), in analogy
to the perturbative pion-pion production (see Eq.(\ref{pQCD_regulator})).
For completeness in the same figure we also show
the cross section for the
$\gamma \gamma \rightarrow \rho^+(0) \rho^-(0)$,
calculated within rules of perturbative QCD \cite{BL81}.

In order to calculate the contributions of $S_2 \pi$ and $H_2 \pi$
to the $\gamma \gamma \rightarrow \pi^+ \pi^-$ reaction
we couple both $\rho^0 \rho^0$ and
$\rho^+(\lambda=0) \rho^-(\lambda=0)$ channels with the
$\pi^+ \pi^-$ channel of interest via standard pion exchange
mechanism. The standard coupling constants obtained from the decay
width of $\rho^0 \rightarrow \pi^+ \pi^-$ are used.
We have used exponential form factors
with $B$ = 4 GeV$^{-2}$ in the fully hadronic vertices.

\begin{figure}
\includegraphics[width=9cm]{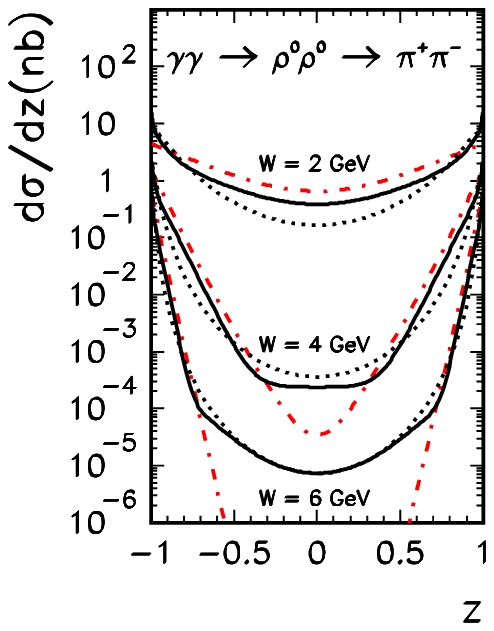}
\caption{
Angular distribution of pions obtained when only the diagram $S_2 \pi$
is taken (dashed). The pion exchange contribution (dotted)
is shown for comparison. The solid line corresponds to a coherent sum
of the pion exchange and $S_2 \pi$ amplitudes.
\label{fig_gg_rhorho_pipi_soft}}
\end{figure}

As in the previous section, we shall quantify the role of
the double-scattering amplitudes.
In Fig.\ref{fig_gg_rhorho_pipi_soft} we show the resulting
angular distributions (dash-dotted lines).
As can be seen from the figure the two-step process
efficiently produces pions at $\theta \sim \pi/2$.
The pion exchange (dotted line) and the coherent sum of pion
exchange and the double scattering $S_2 \pi$ term (solid line)
are shown for comparison. 
The size of the amplitude and the interference with
the pion exchange amplitude depends on photon-photon energy
and $\cos \theta$. For low energy $W \sim$ 2 GeV we observe
\begin{equation}
|\pi|^2 < |\pi + S_2 \pi|^2 < |S_2 \pi|^2 \; .
\end{equation}
At higher energies and $z \sim 0$
\begin{equation}
|S_2 \pi|^2 < |\pi +S_2 \pi|^2 < |\pi|^2 \; .
\end{equation}
This means that in the region of interest, we observe a destructive
interference of the pion exchange and the double-scattering amplitude
$S_2 \pi$. The effect of the $S_2 \pi$ rescattering is rather large.
The situation is, however, not completely clear
in the light of rather small cross section for
$\gamma \gamma \rightarrow \rho^0 \rho^0$ as obtained by
the L3 collaboration \cite{L3_4pi}.
This is a very interesting issue which needs further clarification
but goes beyond the scope of the present discussion.

\begin{figure}
\includegraphics[width=9cm]{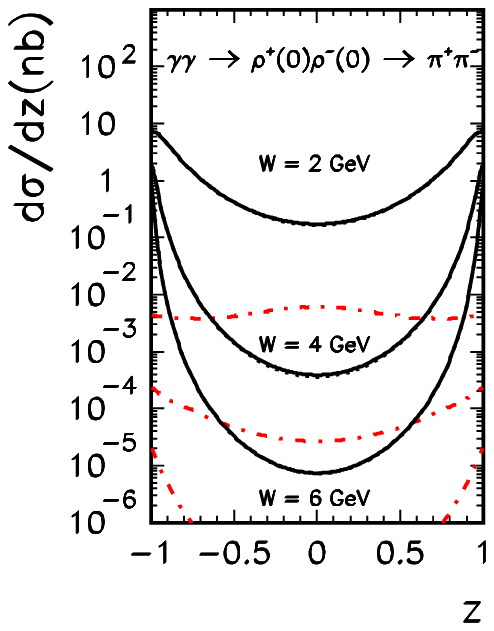}
\caption{
Angular distribution of pions obtained when only the diagram
$H_2 \pi$ is included (dashed). The pion exchange contribution
(dotted) is shown for comparison. The solid line corresponds
to a coherent sum of the pion exchange and $H_2 \pi$ amplitudes.
\label{fig_gg_rhorho_pipi_hard}}
\end{figure}

In Fig.\ref{fig_gg_rhorho_pipi_hard} we show the resulting
angular distributions (dashed lines). The pion exchange (dotted line)
is shown for reference. Due to the helicity structure of the vertices,
the interference between pion-exchange
and the $H_2 \pi$ vanishes identically.
Summarizing, we have shown that the pion-exchange mechanism leads
to a coupling of the $\rho \rho$ and $\pi \pi$ channels.
The interference term of
the pQCD amplitude and the $H_2 \pi$ amplitude was
found to be negative
and its size decreases with photon-photon energy.
The situation for the cross section integrated in the
experimenatal range -0.6 $< z <$ 0.6 can be summarized as follows
\begin{eqnarray}
|H_2 \pi|^2 \ll |pQCD|^2 \; ;
\nonumber \\
|H_2 \pi|^2 < |\pi|^2 \; .
\end{eqnarray}
%


  
Generally the double-scattering effects considered lead to
a mild reduction of the single-scattering amplitudes.
The triple-scattering effects are expected to partially
compensate the double-scattering effects.

\section{All processes together}

\begin{figure}
\includegraphics[width=9cm]{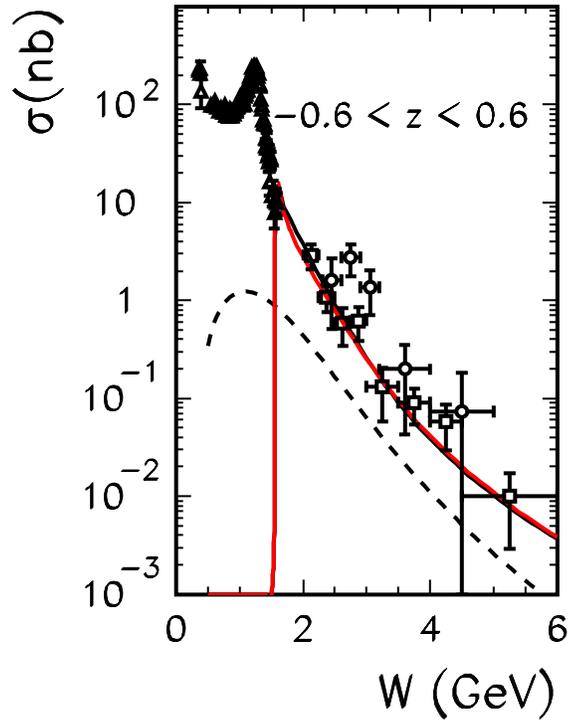}
\caption{
The result corresponding to the coherent sum of all processes
(solid line) versus pQCD result (dashed line).
The vertical solid line shows
the expected lower limit of the range of applicability
of the present multipole-scattering approach.
\label{fig_sum_w}}
\end{figure}

In Fig. \ref{fig_sum_w} we present the result corresponding to
the coherent sum of all processes discussed in the present paper.
For comparison we show the pQCD result.
For W $>$ 2 GeV the final result is practically independent of
the phase of the resonance contribution with repect to
the other contributions.
As can be seen by
comparison of the solid and dashed line the inclusion of
the processes considered in the present paper leads to a
considerable improvement with respect to the pQCD calculation.
The final result describes
the ALEPH \cite{ALEPH} and DELPHI \cite{DELPHI} data
surprisingly well. 

\begin{figure}
\includegraphics[width=9cm]{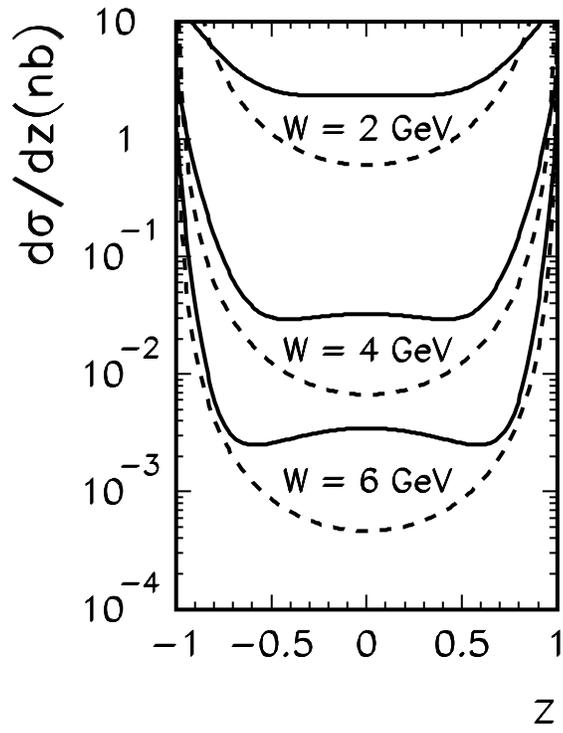}
\caption{
The result corresponding to the coherent sum of all processes
(solid line) versus pQCD result (dashed line).
\label{fig_sum_z}}
\end{figure}

To complete our results in Fig. \ref{fig_sum_z} we
present the corresponding angular distributions.
A large enhancement with respect to pQCD at
intermediate angles (z $\sim$ 0) is clearly visible.
We predict almost flat
$d \sigma / d z$ in the experimentally measured region of
-0.6 $< z <$ 0.6. This is due to the competition of
pion-exchange, pQCD, tail of the $f_2$ resonance
and multiple scattering effects. The contribution of $f_2$
far from the peak position (i.e. for W $>$ 3 GeV) is the least
reliable element of our model calculation. 
In the present analysis we have ignored $\rho$, $a_1$ and $a_2$
meson (reggeon) exchanges which may, at least potentially,
enhance the cross section for $|z| >$ 0.4 - 0.5.
The knowledge of precise angular distributions at different
well defined energies would be helpful to disentangle
the rather complicated reaction dynamics.

\section{Predictions for $\gamma \gamma \rightarrow \pi^0 \pi^0$}

\begin{figure}
\includegraphics[width=9cm]{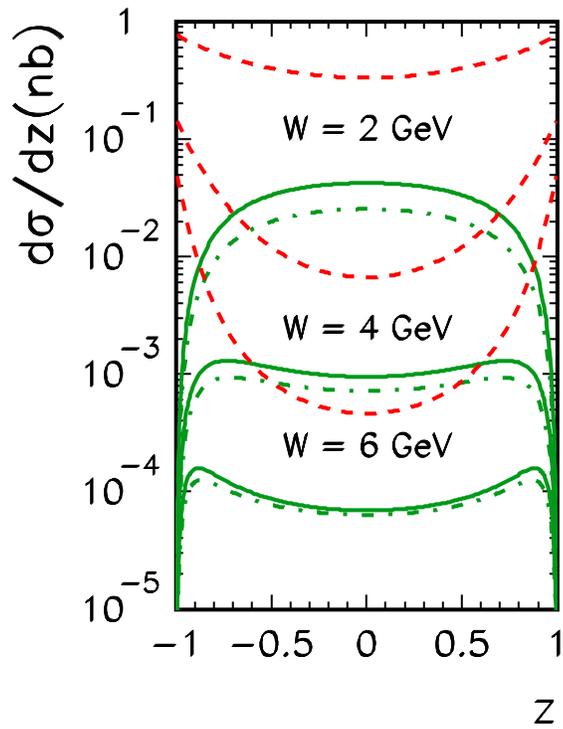}
\caption{
Predictions of pQCD for
angular distributions of the $\gamma \gamma \rightarrow \pi^0 \pi^0$
reaction (solid lines) for W = 2, 4, 6 GeV. 
The solid lines correspond to calculations with running $\alpha_s$
whereas the dash-dotted lines to those with fixed $\alpha_s$.
For reference we show analogous results for the
$\gamma \gamma \rightarrow \pi^+ \pi^-$ reaction (dashed lines).
In all these calculations $\Lambda_{reg}$ = 1 GeV was taken.
\label{fig_n_pqcd}}
\end{figure}

In the case of the $\pi^0 \pi^0$ channel the pion exchange
vanishes identically while the pQCD cross section is rather small.
In Fig.\ref{fig_n_pqcd} we present the predictions of pQCD
for the $\gamma \gamma \rightarrow \pi^0 \pi^0$ reaction
obtained with asymptotic distribution amplitude.
In the case of fixed $\alpha_s$ there are exact cancellations of
different terms in the reaction amplitude for 
$\gamma \gamma \rightarrow \pi^0 \pi^0$ (see for instance \cite{BL81}).
Therefore in the present paper we have performed calculations with
both fixed (dash-dotted lines) and running $\alpha_s$
as proposed in \cite{JA90} (solid lines).
In addition the region of small $t$ and $u$ was
cut off as was the case for the
$\gamma \gamma \rightarrow \pi^+ \pi^-$ reaction
using form factor (\ref{pQCD_regulator}) with $\Lambda_{reg}$ = 1 GeV.
The pQCD cross section for the $\pi^0 \pi^0$ channel
(solid or dash-dotted lines)
is almost an order of magnitude smaller than the pQCD cross section
for the $\pi^+ \pi^-$ channel (dashed lines).

Because of the smallness of the pQCD contribution in the following
we shall consider also other contributions
to the $\pi^0 \pi^0$ channel like $f_2$ resonance contribution
and/or FSI effects due to couplings to other channels.
In calculating the resonance contribution for the
$\gamma \gamma \rightarrow \pi^0 \pi^0$ reaction one should
remember that
\begin{equation}
BR(f_2 \rightarrow \pi^0 \pi^0) = \frac{1}{2} \cdot
BR(f_2 \rightarrow \pi^+ \pi^-) \; .
\end{equation}
Because the FSI effects are relatively weak,
it seems that even far from the $f_2$ peak
the cross section for the $\gamma \gamma \rightarrow \pi^0 \pi^0$
reaction may be dominated by the tail of the broad $f_2$ resonance.  

\begin{figure}
\includegraphics[width=9cm]{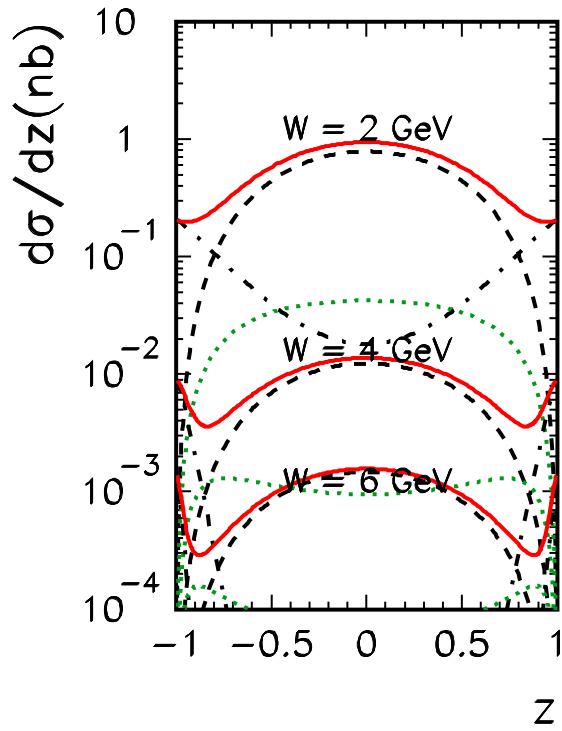}
\caption{
Angular distribution for $\gamma \gamma \rightarrow \pi^0 \pi^0$.
The dash-dotted line is the cross section corresponding to double
scattering contribution (soft-pion-exchange and isovector reggeon).
The dashed line is the $f_2$ contribution and the solid line
corresponds to their coherent sum. For completeness the pQCD
predictions (dotted lines) are shown too.
\label{fig_nsum_z}}
\end{figure}

In Fig.\ref{fig_nsum_z} we present the result
for $\frac{d\sigma}{dz}(\gamma \gamma \rightarrow \pi^0 \pi^0)$
coresponding to a coherent sum (solid line) of the $f_2$ resonance
(dashed line), the pQCD amplitude (dotted line) and
the two-step pion exchange:
$\gamma \gamma \rightarrow \pi^+ \pi^-$
followed by the charge-exchange (CEX) process
$\pi^+ \pi^- \rightarrow \pi^0 \pi^0$ (dash-dotted line).
The parameters of the charge exchange reaction are taken
from \cite{SNS2001}.
One can observe a clear modification of the resonance contribution only
for z $\sim$ 1 and z $\sim$ -1. It would be interesting to extend
the experimentally accessible range of $z$ in the future analyses
in order to identify such effects.
With the phase of the resonance term fixed for the
$\gamma \gamma \rightarrow \pi^+ \pi^-$ reaction other two-step
contributions like hard $\gamma \gamma \rightarrow \pi^+ \pi^+$
followed by the CEX process or
hard $\gamma \gamma \rightarrow \rho^+ \rho^-$ followed by the pion
exchange lead to the negative interference effect
of the order of 10 \% at $z \sim$ 0.

Let us define the quantity:
\begin{equation}
R_{\pi^0 \pi^0 / \pi^+ \pi^-} =
\frac{d \sigma}{dz}(\gamma \gamma \rightarrow \pi^0 \pi^0) \; / \;
\frac{d \sigma}{dz}(\gamma \gamma \rightarrow \pi^+ \pi^-) 
\end{equation}
as a function of $z = \cos\theta$.
Our approach  predicts that $R_{\pi^0 \pi^0 / \pi^+ \pi^-}$
strongly depends on $z$.
For z $\sim$ 0 our result is in between that predicted by pQCD
$R_{\pi^0 \pi^0 / \pi^+ \pi^-} \approx$ 0 
and that from the recent approach \cite{DKV02} where   
$R_{\pi^0 \pi^0 / \pi^+ \pi^-}$ = 1 has been predicted.
In principle this significant difference between these all
three approaches should be easy to verify experimentally.

\newpage

\section{Conclusions}

For a long time the $\gamma \gamma \rightarrow \pi^+ \pi^-$ reaction
was considered as a gold-plated reaction to identify
pQCD effects.
Recent LEP2 results for $\gamma \gamma \rightarrow \pi^+ \pi^-$,
when combined with the present understanding of the pion
distribution amplitudes, clearly demonstrate a failure of leading
twist pQCD in explaining the data.
In this analysis we have thoroughly studied the role of different
mechanisms which could potentially cloud the pQCD contribution:
\begin{itemize}
\item soft pion exchange
\item tail of the broad $f_2$ resonance
\item interference of pion-exchange and pQCD amplitudes
\item multipole scattering of pions
\item coupling between $\rho \rho$ and $\pi^+ \pi^-$ channels
\end{itemize}

It was found that the soft pion exchange, which is known to be
the dominant mechanism below the $f_2$ resonance, stays
important also above the resonance.
More subtle details are however difficult to predict as the
half-off shell electromagnetic form factor of the pion
is rather poorly known both experimentally and theoretically.
In addition the pion-exchange amplitude strongly interferes
with the pQCD amplitude.

Because the pQCD contribution strongly decreases with
the photon-photon energy the high energy flank of the broad
$f_2$ resonance gives a contribution of comparable size
to the pQCD continuum in a rather broad energy range
above the resonance.

Using phenomenological, yet realistic, pion-pion interaction
found in a recent analysis we have estimated the contribution
of the final state interaction processes.
We have found that pion-pion FSI as well as the coupling
with the $\rho \rho$ channel modify the total amplitude
at the level of 10 - 20 \%.

In our model we predict: 
$\frac{d \sigma}{dz}(\gamma \gamma \rightarrow \pi^0 \pi^0) < 
 \frac{d \sigma}{dz}(\gamma \gamma \rightarrow \pi^+ \pi^-)$.
The leading mechansims for the
$\gamma \gamma \rightarrow \pi^0 \pi^0$ being $f_2$
resonance and FSI rescattering effects, including coupling
with the $\rho \rho$ channel.
This result differs from the recent predictions
of Diehl, Kroll and Vogt \cite{DKV02} who obtained:
$\frac{d \sigma}{dz}(\gamma \gamma \rightarrow \pi^0 \pi^0) \approx 
 \frac{d \sigma}{dz}(\gamma \gamma \rightarrow \pi^+ \pi^-)$.
We hope that experimental verification of both scenarios
will be possible in near future.
In our opinion the effects discussed in the present paper
must be taken into account in order to extract empirically
the hand-bag contribution. This may not be an easy task,
however, as the different contributions can interfere.
Good data for both $\pi^+ \pi^-$ and $\pi^0 \pi^0$
channels, including angular distributions, also at
$|\cos\theta| >$ 0.6, would be very helpful to disentangle
the complicated dynamics in the region of intermediate energies.

In the light of our analysis we conclude that the
$\gamma \gamma \rightarrow \pi^+ \pi^-$ reaction is
not the best choice to identify the pQCD predictions.
We plan a similar analysis for other meson pair production
in $\gamma \gamma$ collisions in order to find the best
candidate to study perturbative QCD.
The $\gamma \gamma \rightarrow K^+ K^-$ reaction seems to be a
better candidate.

\vskip 1cm

{\bf Acknowledgments}

A.S. thanks Siggi Krewald for interesting discussion
on intermediate energy physics and Kolya Nikolaev
for the discussion of pion-pion multipole scattering effects.
The discussion with Carsten Vogt on his improved pQCD
predictions \cite{Vogt} is gratefully acknowleged.
We are indebted to Katarzyna Grzelak and Alex Finch
for providing us with the DELPHI and ALEPH experimental data
for $\gamma \gamma \rightarrow \pi^+ \pi^-$ shown at the
PHOTON2001 workshop \cite{DELPHI,ALEPH} and discussion.
One of us (A.S.) is indebted to Nikolai Achasov for
drawing our attention to experimental data on
$\gamma \gamma \rightarrow \rho^0 \rho^0$.
This work was partially supported by the German-Polish DLR exchange
program, grant number POL-028-98.


\newpage


\begin{figure}
\includegraphics[width=9cm]{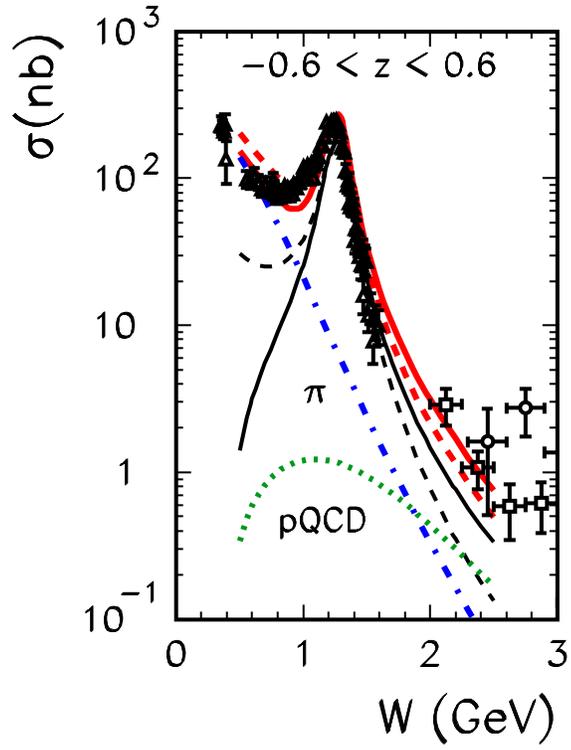}
\caption{
  Interference effect of resonant and nonresonant contributions. The
  thick lines represent the sum of pion-exchange, pQCD and $f_2$
  contributions with energy dependent (solid line) and energy
  independent (dashed line) width of the resonance.  For reference we
  display the pion-exchange (dash-dotted) and pQCD (dotted line)
  contributions.
\label{fig_f2_interference}}
\end{figure}


\end{document}